\documentclass[12pt]{iopart}

\usepackage{graphicx}
\usepackage{cite}
\usepackage{color}
\usepackage{hyperref}

\begin{document}

\title[1D disordered Bose-Hubbard model: superfluid density \& quasi-long-range order]{
Superfluid density and quasi-long-range order in the one-dimensional disordered Bose-Hubbard model}

\author{M. Gerster$^1$, M. Rizzi$^2$, F. Tschirsich$^1$, P. Silvi$^1$, R. Fazio$^{3,4}$ and S. Montangero$^{1,5}$}
\address{$^1$ Institute for Complex Quantum Systems, Ulm University, D-89069 Ulm, Germany}
\address{$^2$ Johannes Gutenberg-Universit\"at Mainz, Institut f\"ur Physik, Staudingerweg 7, D-55099 Mainz, Germany}
\address{$^3$ ICTP, Strada Costiera 11, I-34151 Trieste, Italy}
\address{$^4$ NEST, Scuola Normale Superiore \& Istituto Nanoscienze CNR, I-56126 Pisa, Italy}
\address{$^5$ Center for Integrated Quantum Science and Technology (IQST), Universities of Ulm/Stuttgart and MPI for Solid State Research, Germany}
\ead{matthias.gerster@uni-ulm.de}

\begin{abstract}
We study the equilibrium properties of  the one-dimensional disordered Bose-Hubbard model by means of a  
gauge-adaptive tree tensor network variational method suitable for systems with periodic boundary conditions. We
compute the superfluid stiffness and superfluid correlations close to the superfluid to glass transition line, 
obtaining accurate locations of the critical points. By studying the statistics of the exponent of the 
power-law decay of the correlation, we determine the boundary between the superfluid region and 
the Bose glass phase in the regime of strong disorder and in the weakly interacting region, 
not explored numerically before. In the former case our simulations are in agreement with 
previous Monte Carlo calculations.
\end{abstract}



\maketitle

\section{Introduction}

The Bose-Hubbard model is a widespread paradigm in the field of strongly-correlated 
many-body systems~\cite{Fisher1989}. Over the years it has found applications in the physics of superconducting 
thin-films~\cite{Fisher1990}, Josephson junction arrays~\cite{Fazio2001}, 
optical cavity arrays~\cite{Tomadin2010}, and cold atoms in optical lattices~\cite{Jaksch1998}.
The model describes bosons hopping on a tight-binding lattice and interacting through an on-site repulsion. 
The non-trivial properties of the model stem from the 
competition between the hopping (that delocalizes the particles) and the local repulsion (that 
tends to suppress local particle number fluctuations). 
In the absence of disorder such a competition dictates a quantum phase transition at integer filling:
if the hopping dominates, the ground state of the system is superfluid;
in the opposite regime, there is a gap in the excitation spectrum and the system is in a Mott insulating state.
At a critical value of the ratio between the hopping and the repulsion parameters,  
an insulator to superfluid quantum phase transition takes place at zero temperature.
In optical lattices this transition has been experimentally demonstrated in the seminal work by Greiner 
{\it et al.}~\cite{Greiner2002}. The phase diagram of the Bose-Hubbard model has been studied 
by a number of analytical and numerical methods, a recent account of the properties of this 
model together with a review of the experimental advances with cold atoms can be found in the 
reviews~\cite{Lewenstein2007,Bloch2008}.  

In one dimension (1D), which is the case of interest for the present work, the quantum phase transition 
between the Mott insulating (MI) phase and the superfluid (SF) phase  at fixed commensurate
lattice filling is known to be of the Berezinkii-Kosterlitz-Thouless (BKT) type~\cite{Fisher1989}. 
The critical properties of the quantum model map onto that of a classical 
two-dimensional $XY$ model.
The hallmark of a BKT transition~\cite{Berezinskii1971,Kosterlitz1973,Kosterlitz1974} is a universal 
jump~\cite{Nelson1977} in the superfluid density (or stiffness or helicity modulus~\cite{Weber1988}).  
Extensive studies of the stiffness in the classical 2D $XY$ model are reported 
for example in~\cite{Weber1988,Hsieh2013}. Monte Carlo simulations of the one-dimensional 
quantum model (the mapping onto the $XY$ model becomes  accurate at large fillings) were 
performed originally in~\cite{Batrouni1990}, more recent results are comprehensively discussed 
in~\cite{Cazalilla2011}.
On the other hand, the Density Matrix Renormalization Group alias 
Matrix Product States (MPS)~\cite{Schollwoeck2005-2011}
has been applied to the model under consideration~\cite{Kuehner2000,Rizzi2005,Doria2011,Rossini2012}, 
although the superfluid stiffness is difficult to extract with high precision 
because these methods are not ideally suited to deal with 
periodic boundary conditions (PBC)~\cite{periodic}. 

The presence of disorder enriches the phase diagram with a new phase, the Bose glass (BG)~\cite{Fisher1989}, 
surrounding a superfluid lobe in the interaction vs. disorder plane of the phase diagram. 
In absence of interactions, indeed, the disorder induces Anderson localization of single particle states 
and therefore an insulating phase. This is then lifted by a moderate amount of interactions that tend
to reestablish coherence and consequently superfluidity.
At large interactions and large disorder, finally, strong correlations should lead back to an insulating phase,
distinct from the above mentioned Mott insulator in the clean setup.
Although this picture is qualitatively well-understood, the full quantitative characterization of the 
disordered Bose-Hubbard model remains challenging:
considerable efforts have been devoted to study its phase diagram, 
in particular the location and the nature of the 
glassy phase~\cite{Giamarchi1988,Pai1996,Herbut1998,Prokofev1998,Rapsch1999,Pollet2009,Carrasquilla2010, 
Altman2004,Pielawa2013,Pollet2013a,Ristivojevic2014,Goldsborough2015a}.
Developments on the issue, with a special emphasis on numerical simulations, 
have been reviewed in~\cite{Pollet2013b}. 
Recently, an experiment with optical lattices has made an important step forward 
in mapping the phase diagram from weak to strong interactions~\cite{Derrico2014}.

In the present work we apply a Tree Tensor Network (TTN) 
method~\cite{Shi2006,Murg2010,Silvi2010,Goldsborough2014,Goldsborough2015b} 
(specifically our recently introduced gauge-adaptive TTN technique~\cite{Gerster2014}), 
a natural ansatz for the simulation of one-dimensional quantum many-body systems with PBC,
to the disordered Bose-Hubbard model. We compute the superfluid stiffness and correlation functions, 
avoiding the detrimental boundary effects arising in simulations with open boundary conditions.  
Furthermore, by analyzing the 
statistics of the decay of superfluid correlators we are able to locate the superfluid to Bose glass 
transition line, to extend it into the previously numerically unexplored regime of small on-site 
interactions, and to confirm the Monte Carlo results of Ref.~\cite{Prokofev1998} for strong interactions.

The paper is organized as follows: In the next section we introduce the model and give a brief overview of the 
used numerical technique (a detailed discussion can be found in~\cite{Gerster2014}) 
and we anticipate the main result of the paper, the phase diagram of the disordered BH model at unit filling. In
section~\ref{sssection} we analyze the superfluid stiffness. As a benchmark we first consider the case without
disorder where we are able to perform a very accurate finite-size scaling to locate the BKT transition. We compare the obtained 
transition point with the results from other well-known methods for determining the MI-SF transition, 
such as numerically extracting the Luttinger parameter from the 
hopping correlation functions \cite{Kuehner2000}, finding very good agreement. Moreover, we present our results
for the average stiffness in the disordered case.
In order to get a precise location of the transition points we study the hopping correlation decay, as presented  
in section~\ref{corrsection}. A detailed analysis of the statistics of the exponent of the power-law decay allows us 
to map out the complete SF-BG transition line in the phase diagram. 
Finally, we summarize the main results of the work in section~\ref{conclusions}.

\section{The disordered Bose-Hubbard model}
\label{BHsection}

The one-dimensional Bose-Hubbard model is a system of spinless bosons moving in a single-band tight-binding
lattice, described by the Hamiltonian
\begin{equation}
		{\cal H} = -t \sum_j \left( b_j^\dagger b_{j+1} + \mathrm{h.c.} \right)
		+ \frac{U}{2} \sum_j n_j (n_j - 1) -
		\sum_j V_j n_j \; .
\label{eq:ham_bhring}
\end{equation}
The first term accounts for the hopping between neighboring sites (labeled by the  index~$j$), the second for
the on-site repulsion and the last for the effect of a random inhomogeneous chemical potential. 
The couplings $t$, $U$ and $V_j$ are the hopping amplitude, 
the on-site repulsion, and the random potential, respectively;
$b_j$ ($b^\dagger_j$) are bosonic annihilation (creation) operators obeying $[b_j, b^\dagger_{j\prime}]=
\delta_{jj^\prime}$, with $n_j = b^\dagger_j b_j$ being the corresponding number operators. 
The random on-site disorder $V_j$ is independent on different sites and uniformly distributed in 
the interval $[-\Delta, \Delta]$. 

\begin{figure}
	\centering
	\includegraphics{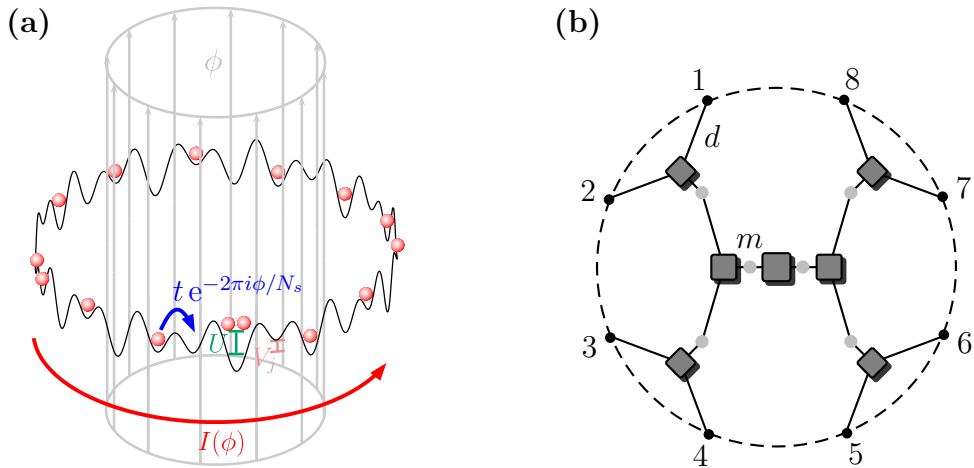}
	\caption{\label{fig:bhring_sketch}(a) Sketch of a Bose-Hubbard ring with hopping $t$, 
	on-site interaction $U$ and local disorder $V_j$ 
	pierced by an external flux $\phi$, giving rise to a 
	ground state boson current $I(\phi) \propto \partial E_0 / \partial \phi$.
	(b) Sketch of the TTN state for a lattice with
	$N_s=8$ sites. The lattice sites (black dots) have a local dimension $d$. The
	three-legged tensors (gray squares, each leg represents a tensor index) 
	hierarchically group two sites together, forming virtual sites (light gray dots).  
	The dimension of the Hilbert space of the virtual sites is truncated at some
	maximum bond dimension $m$. 
	A variational optimization of this RG scheme is then performed following 
	the prescriptions introduced in~\cite{Gerster2014}.
	}
\end{figure}

In order to compute the stiffness we introduce a twist $\phi$ in the boundary conditions. This amounts to 
considering periodic boundary conditions and performing a Peierls shift in every hopping matrix element
$$
  t \, \left( b_j^\dagger b_{j+1} + \mathrm{h.c.} \right) 
  \;\; \longrightarrow \;\; 
  t \, \left( b_j^\dagger b_{j+1} \; \mathrm{e}^{-2\pi i \phi / N_s} + \mathrm{h.c.} \right) 
$$
where $N_s$ is the number of sites of the ring, as schematically depicted 
in Fig.~\ref{fig:bhring_sketch}~a) where the 
twist is explicitly interpreted as a magnetic flux $\phi$ piercing the ring and acting on 
electrically charged particles. 

Our TTN ansatz is sketched in Fig.~\ref{fig:bhring_sketch}~b), showing the natural
capability to represent states on 1D lattices with PBC, despite its loop-free
geometry. The absence of loops in the network is crucial for the applicability of
efficient energy minimization techniques needed for the ground state search.
For algorithm details we refer the reader to Ref.~\cite{Gerster2014}, however  
it is important to emphasize that with our TTN approach we will be able to compute the superfluid 
density for large system sizes (up to $N_s=256$ in the absence of disorder) 
and therefore perform a reliable finite-size scaling analysis. As we will show, 
at equilibrium in one dimension this approach provides matching results to Monte Carlo calculations. 

The TTN algorithm we use 
has abelian global symmetries built-in~\cite{Singh2010}, 
allowing us to work in a canonical ensemble (at zero temperature)
by targeting exclusively $N$-particles states, in accordance with the $U(1)$ particle conservation symmetry
$[{\cal H}, \sum_j n_j]=0$ of the Hamiltonian ${\cal H}$.  In the rest of the paper we will in particular 
consider the case of unit filling, i.e. $N=N_s$. We simulate system sizes of 
up to $N_s=256$ sites, with a truncated local boson basis up to a dimension of $d=20$ in
the low $U$ regime where high local populations are possible. In the disordered scenario   
we average all observables over a number of $n_r=100$ different disorder realizations.

\begin{figure}
	\centering
	\includegraphics{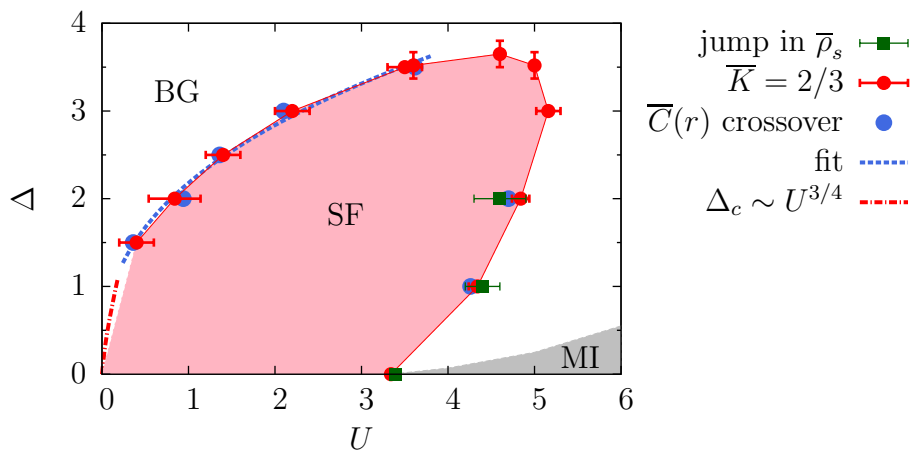}
	\caption{\label{fig:pd_dirtybosons} Phase diagram of the BH model
	with disorder in 1D at unit filling: Bose glass (white region), superfluid lobe (red region), 
	Mott insulator (gray region). The boundaries of the SF lobe are determined via three different criteria:
	superfluid jump (green squares), Luttinger parameter $\overline{K}=2/3$ (red circles) and 
	algebraic versus exponential decay of correlation functions (blue circles). 
	The blue dashed line is a fit through the blue circles according to $\mu \, U^\nu$, 
	with fit parameters $\mu=2.2 \pm 0.1$ and $\nu=0.4 \pm 0.1$
	The red dashed line follows the conjectured behavior $\Delta_c \sim U^{3/4}$ of
	the phase boundary for $U \ll 1$~\cite{Pollet2013b}. The gray shaded MI region
	is data from Ref.~\cite{Prokofev1998}.
	}
\end{figure}

In Fig.~\ref{fig:pd_dirtybosons} we anticipate the main result of this work, 
the phase diagram of the disordered BH model. The technical details and the discussion of the results 
are presented in the following sections, here we highlight that for 
strong interactions (the right edge of the lobe) our results obtained with 
three different criteria are in agreement with state-of-the-art Monte Carlo results. 
More interestingly, we are able to extend our numerical analysis into the 
low-interaction regime (left edge of the lobe): two independent criteria strongly 
agree down to values of the interactions of the order of $U \sim 0.4$, 
where the numerical results nicely match the theoretical conjecture for very small $U$ \cite{Pollet2013b}.

\section{Superfluid stiffness}
\label{sssection}

The superfluid stiffness $\rho_s$ is defined through the sensitivity of the ground state 
energy~$E_0$ (here we consider the zero-temperature case) to a twist in the boundary conditions
\begin{equation}
	\rho_s =  \frac{N_s}{8\pi^2} \frac{\partial^2 E_0}{\partial \phi^2}\bigg|_{\phi=0} \; , 
\label{eq:superfluid_density}
\end{equation}
and we set the energy scale by choosing $t=1$. Therefore, $\rho_s = 1$ for vanishing on-site repulsion~$U$
and vanishing disorder $\Delta$, while $\rho_s = 0$ in the Mott-insulator and Bose glass phases.

\begin{figure}
	\centering
	\includegraphics{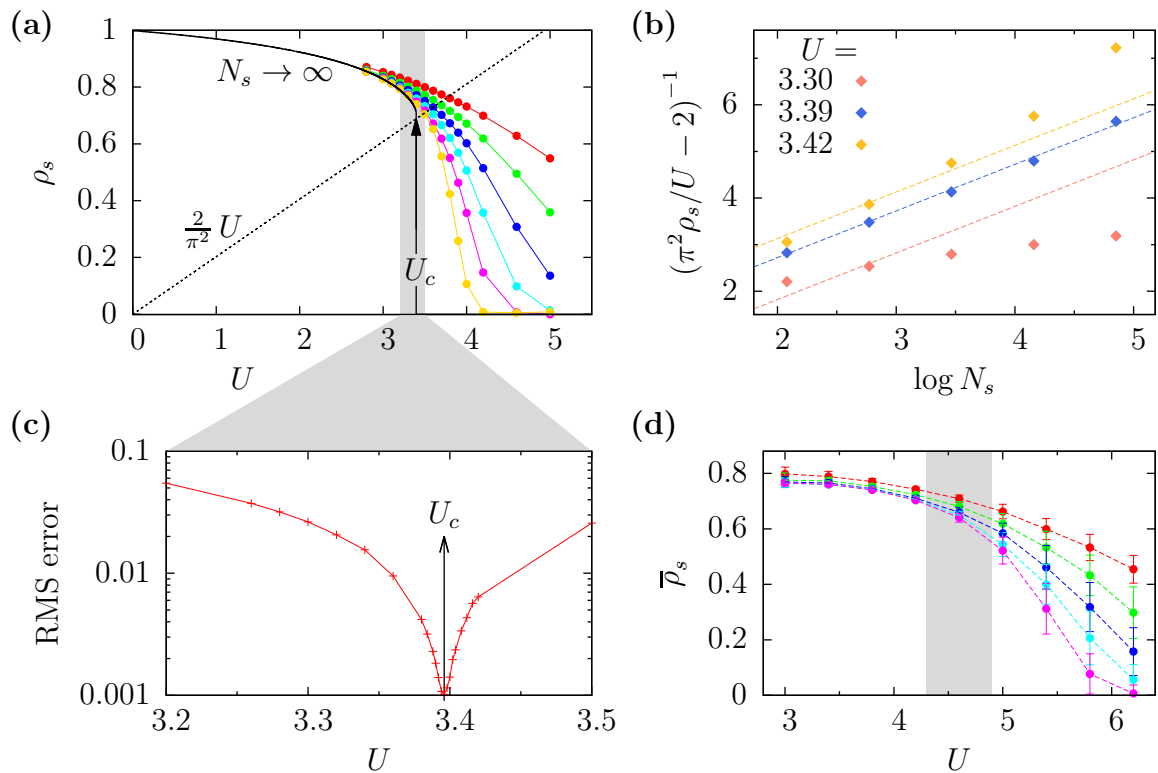}
	\caption{(a) Superfluid density $\rho_s$ as a function of
	on-site repulsion $U$ for various ring sizes $N_s=8,16,32,\ldots,256$ (from top to bottom) 
	in the clean case $\Delta=0$. The arrow indicates the 
	location and the height of the jump of $\rho_s$ in the thermodynamic limit, 
	determined via Weber-Minnhagen fitting (shown in panel~(b), dashed lines are fits). 
	The resulting RMS errors for the gray-shaded interval of $U$ are shown in 
	panel~(c), where the pronounced minimum at $U_c = 3.394 \pm 0.03$ becomes apparent.
	(d) Averaged superfluid density	$\overline{\rho}_s$ as a function of 
	on-site repulsion $U$ for $\Delta=2.0$ in the disordered case. 
	The error bars indicate the standard deviations of the $\rho_s$ distributions. 
	The gray-shaded interval is the location of the critical point $U_c = 4.6 \pm 0.3$, 
	determined via Weber-Minnhagen fitting.
	}
\label{fig:supfluiddens}
\end{figure}

As a benchmark of our approach, we first consider the clean case $\Delta = 0$. At the 
BKT critical point the superfluid density jumps from a finite value $\rho^c_s$ to zero. The 
mapping onto the classical $XY$ model yields the critical value 
$\rho^c_s = \left( 2/\pi^2 \right) U_c$ in  resemblance with the relation
$P^{\,c}_s = \left( 2/\pi \right) T_c$, known to hold 
for the classical XY model in 2D~\cite{Nelson1977}. 
We determine the transition point $U_c$ by extracting the superfluid density 
as a function of the on-site repulsion $U$ for 
different ring sizes $N_s$. We then perform a Weber-Minnhagen fit according to the 
theoretically predicted functional dependence of the superfluid density at 
the BKT transition~\cite{Weber1988}:
\begin{equation}
	\rho_s^c(N_s) = \rho_s^c \left(1 + \frac{1}{2} \frac{1}{\log N_s + b} \right) \; ,
	\label{eq:weber_minnhagen}
\end{equation}
with $b$ the only fit parameter. 
When fitting $\rho_s(N_s) = 2U/\pi^2 \, \left(1 + [2 \log N_s + b]^{-1}\right)$ 
for various $U$, the root-mean-square (RMS) error of the fit displays a pronounced
minimum at the transition point~\cite{Weber1988}.
Figs.~\ref{fig:supfluiddens}~a)--c) show the result of this procedure, resulting 
in~$U_c = 3.394 \pm 0.03$ (or, equivalently, $t_c/U= 0.295 \pm 0.003$).

We now introduce disorder and repeat the study for the averaged superfluid density~$\overline{\rho}_s$.
Since the SF-BG transition is also expected to be of the BKT type~\cite{Giamarchi1988}, 
at least on the right side of the SF lobe (i.e. at large $U$),  
$\overline{\rho}_s$ should display a jump at the phase transition in the thermodynamic limit.
An example of our numerical results is shown in Fig.~\ref{fig:supfluiddens}~d). 
In qualitative terms, for both sides of the lobe a drop of the 
superfluid density for increasing system sizes can be observed when entering the BG phase. 
Like in the clean system, we localize the critical point by means of a 
Weber-Minnhagen fitting procedure, this time with $\overline{\rho}^c_s$ as 
an additional fit parameter because the height of the jump 
for $N_s \rightarrow \infty$ is not known for arbitrary disorder.
While on the right boundary of the SF lobe we are able to successfully apply this technique
employing system sizes and disorder realization numbers accessible to our numerics, 
this is not the case on the left boundary. 
Here we do not achieve sufficient numerical precision to estimate a reliable minimum in the RMS error 
when fitting Eq.~\eref{eq:weber_minnhagen}, 
which ultimately makes this method of determining the phase transition unfeasible for the small $U$ regime.
This might be as well an indication of a different nature of the SF-BG phase transition in this regime,
as proposed in some studies~\cite{Pollet2013b}. 

In order to continue with an accurate analysis of the critical SF-BG line in all regimes we study 
the superfluid correlations in the next section.

\section{Correlation functions }
\label{corrsection}

\begin{figure}
	\centering
	\includegraphics{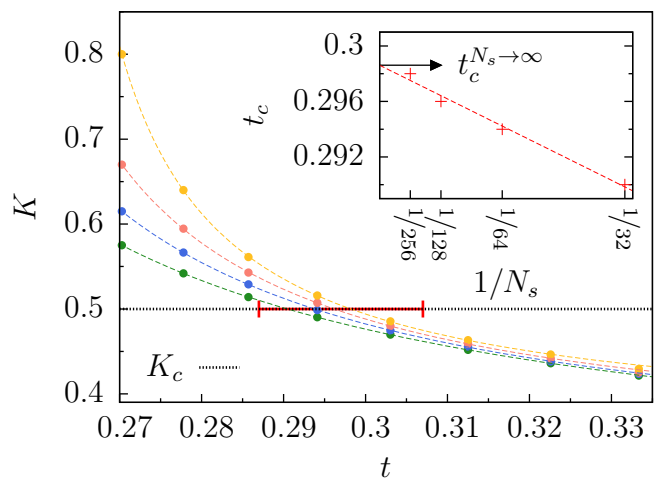}
	\caption{\label{fig:luttparam_finsize}Determination of the clean ($\Delta = 0$) BKT transition point 
	via the Luttinger parameter criterion $K_c=1/2$. The highlighted interval on the
	$K_c$ line is the result obtained in Ref.~\cite{Kuehner2000}. The four different curves correspond to
	the four different system sizes $N_s=32, 64, 128, 256$ (bottom to top). The inset
	shows an extrapolation of $t_c$ to the thermodynamic limit 
	by means of a fit linear in $1/N_s$ (red dashed line), 
	resulting in $t_c^{N_s\rightarrow\infty}=0.299 \pm 0.002$.
	Errors due to finite bond dimension of the TTN state are smaller than the point size (see
	\ref{app:corr} for a convergence check).}
\end{figure}

We proceed further by studying the correlation functions, 
a method that turns out to yield higher accuracy in determining the transition line. 
Moreover, we double-check our results via another method 
for the determination of the MI-SF transition, namely the value of the Luttinger 
parameter $K$ obtained from the hopping correlation functions~\cite{Kuehner2000}
\begin{equation}
	C(r) = \frac{1}{N_s} \sum_j \langle b^\dagger_j b_{j+r} \rangle \propto r^{-K/2} \; .
	\label{eq:hopping_corr}
\end{equation}
In the clean case the criterion for locating the BKT transition is given by $K_c=1/2$.
Indeed, as critical exponents can be obtained with high numerical 
precision in TTN simulations \cite{Gerster2014}, we expect this 
method to work well in the present scenario. Moreover, the PBC setting allows
for an easy treatment of finite size effects in Eq.~\eref{eq:hopping_corr},
simply by replacing the distance $r$ with the effective distance $\tilde{r} =
\mathrm{crd}(r)$, where $\mathrm{crd}(r) = N_s/\pi \, \sin(\pi r / N_s)$ is the
chord function giving the effective distance of two sites $j$ and $j+r$  
arranged on a regular polygon of $N_s$ edges. In \ref{app:corr} we demonstrate the validity 
of this procedure.
Fig.~\ref{fig:luttparam_finsize} summarizes
our numerical results, locating the transition at $t_c = 0.299 \pm 0.002$.
Here, we set $U=1$ for the sake of an easy comparison with
Ref.~\cite{Kuehner2000}; see Tab.~\ref{tab:results_compare} for how this
compares to the transition points obtained in the literature.

We now once again turn back to the study of the disordered case.
As already mentioned, this long-standing problem is unsolved in the low-interaction regime $U<2$,
and even for the more easily tractable regime of higher interactions,
results from different numerical methods are not in
quantitative agreement~\cite{Pai1996, Prokofev1998, Rapsch1999, Pollet2013b}.

\begin{figure}
	\centering
	\includegraphics{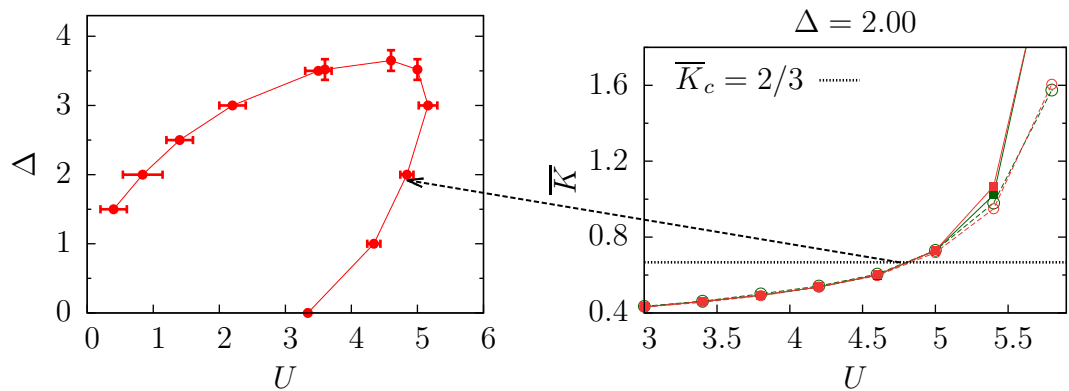}
	\caption{\label{fig:luttparam_criterion}Extension of the SF lobe
	determined from the criterion $\overline{K}_c=2/3$ (left). The right panel
	shows the averaged Luttinger parameter $\overline{K}$ as a function of $U$ at
	fixed disorder $\Delta=2$, for two different ring sizes $N_s=64, 128$ (circles, squares)
	and two different numbers of disorder realizations $n_r=50, 100$ (green, red).
	In the low $U$ regime we used local dimensions of up to $d=20$ in
	order to account for possible high on-site occupations.
	}
\end{figure}

We start by studying the hopping correlation functions of Eq.~\eref{eq:hopping_corr},
whose critical exponent gives access to the Luttinger parameter $K$. 
The Giamarchi-Schulz criterion $\overline{K}_c=2/3$ \cite{Giamarchi1987, Giamarchi1988}, 
where the overline indicates an
average over a number of $n_r$ disorder realizations, results in the phase boundary 
shown in Fig.~\ref{fig:luttparam_criterion}. 
We remark that while the $\overline{K}_c=2/3$ criterion is known to be 
appropriate in the high $U$ regime \cite{Pollet2013b}, 
there is no definite evidence that this is also the case on the left boundary of the SF lobe (low $U$ regime).
Therefore, the results of Fig.~\ref{fig:luttparam_criterion} require critical verification,
which is presented in the following paragraphs.

\begin{figure}
	\centering
	\includegraphics{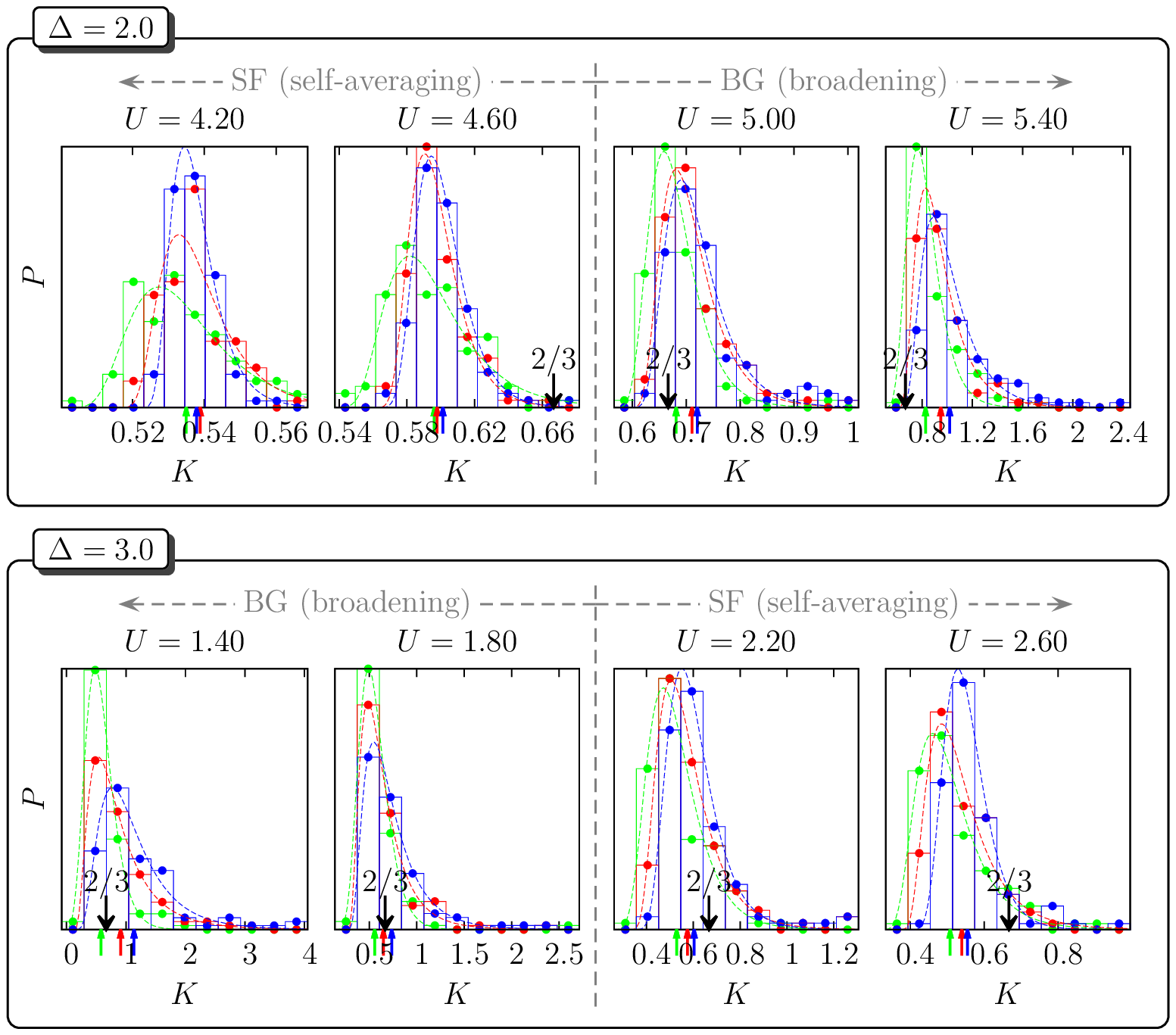}	
	\caption{\label{fig:dist_behavior}Behavior of the distributions of the
	Luttinger parameter $K$ across the SF-BG transition for different system sizes
	$N_s=32, 64, 128$ (green, red, blue), illustrated for two
	disorder strengths $\Delta=2$ (top panel) and $\Delta=3$ (bottom
	panel). Dashed lines are fits according to log-normal distributions 
	Eq.~\eref{eq:lognormal}, colored arrows indicate the corresponding mean values. 
	Note the different scales on the abscissae to
	appreciate the clear difference between the self-averaging and broadening
	regimes. In all plots the number of disorder realizations is $n_r=100$.
	Increasing $n_r$ leaves the properties of the distributions unchanged within the
	statistical error, as we verify in~\ref{app:sample_number}.
	}
\end{figure}

As a first step into this direction we study the distributions of the Luttinger parameter $K$  
over disorder realizations.
It has been shown~\cite{Pollet2013a} that in the SF phase the distributions
display self-averaging, i.e. they become sharper and narrower for
increasing system sizes. This is contrasted by a broadening and flattening in
the BG phase~\cite{Pollet2013a,Zuniga2015}. 
Therefore, the change between the two different types of
behaviors should signal the SF-BG transition. Fig.~\ref{fig:dist_behavior}
shows our analysis for two different values of the disorder strength: the first one contains
the crossing of the right phase boundary when moving along the line $\Delta=2$,
and the second one shows the crossing of the left phase boundary along the line
$\Delta=3$. We find that the $K$-distributions are well captured by log-normal
distributions with probability density
\begin{equation}
   P(x) = \frac{1}{\sigma\sqrt{2\pi}} \, \frac{1}{x} \, 
   \exp\left( -\frac{1}{2} \left(\frac{\log x-\mu}{\sigma} \right)^2 \right) \; .
   \label{eq:lognormal}
\end{equation}
Here $x = K - K_\mathrm{min}$  (where $K_\mathrm{min}$ is the smallest measured
$K$ value among the $n_r$ different disorder realizations), since $P(x < 0) = 0$
for log-normal distributions. 
The parameters $\mu$ and $\sigma$ determine
the mean and the variance of the distribution. The results in 
Fig.~\ref{fig:dist_behavior} are consistent with the results in
Fig.~\ref{fig:luttparam_criterion}, meaning that we observe the crossover from
self-averaging ($\partial[\mathrm{Var}(K)]/\partial N_s < 0$) to 
broadening ($\partial[\mathrm{Var}(K)]/\partial N_s > 0$) at locations 
compatible with the criterion $\overline{K}_c = 2/3$. 

\begin{figure}
	\centering
	\includegraphics{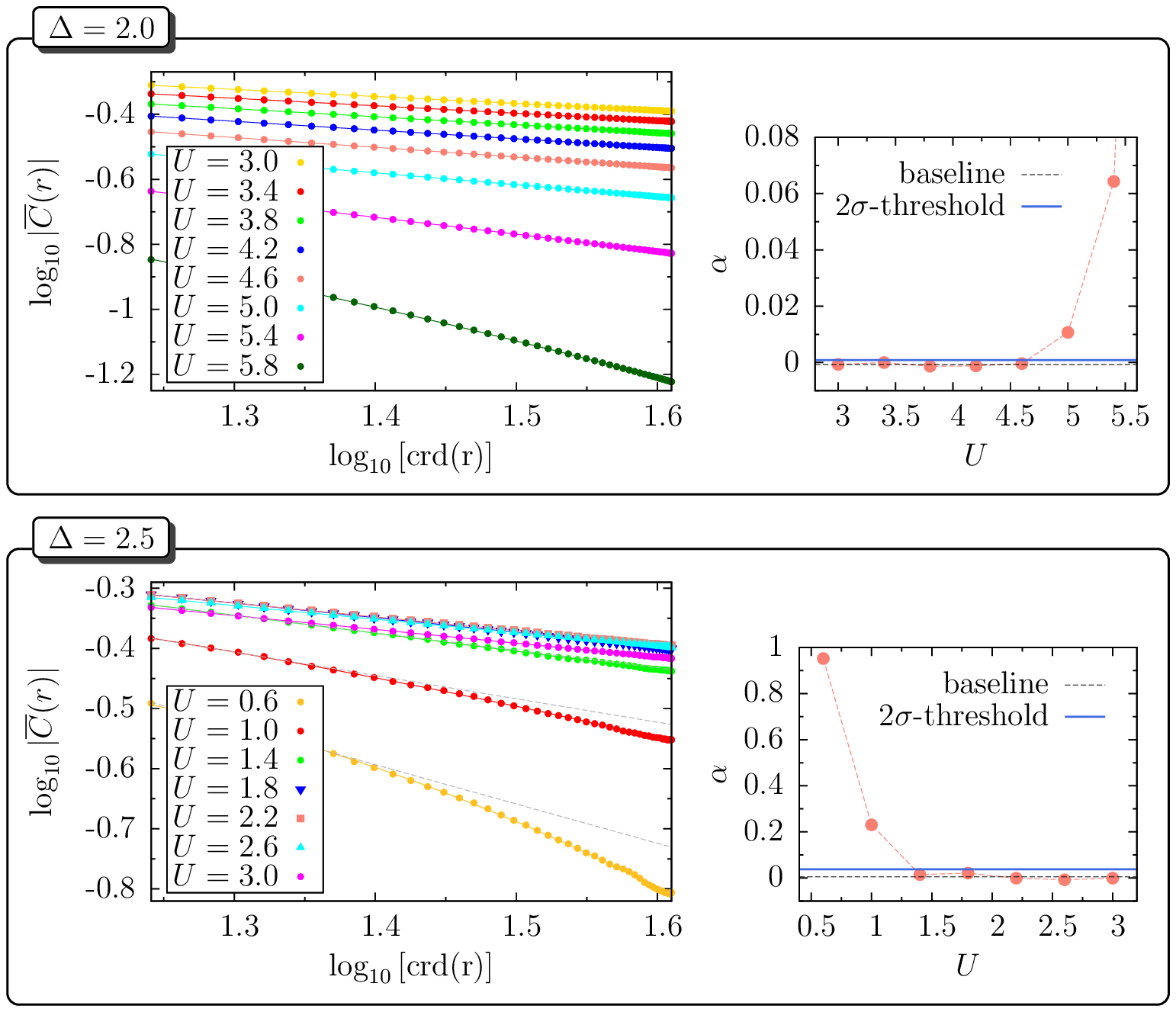}	
	\caption{\label{fig:hoppcorr_averaged}Determining the location of the SF-BG
	transition via the averaged hopping correlation functions.
	The left plots show the averaged correlation functions
	$\overline{C}(r)$ together with fits according to Eq.~\eref{eq:hoppcorr_log}. 
	The dashed gray lines are linear fits through the first
	five data points, serving as guides to the eye to discriminate the different scaling behaviors.
	The system size is $N_s=128$, the number of disorder realizations is $n_r=100$. 
	Increasing $n_r$ does not result in changes bigger than the point size, 
	the same applies for finite bond dimension effects. The extracted curvatures $\alpha$ as a function
	of $U$ are shown in the right plots. The baselines are determined as the
	average of the data points close to zero; the threshold is used to locate
	the critical points.}
\end{figure}

Finally, we propose a method for locating the SF-BG transition that requires minimal
a priori knowledge on the nature of the transition, but merely relies on the
assumption that the averaged hopping correlation functions $\overline{C}(r)$
(understood as an average of the correlation function $C(r)$ in Eq.~\eref{eq:hopping_corr}
over all disorder realizations for each distance $r$) 
decay \emph{algebraically} in the SF phase while in the localized BG phase
they decay \emph{exponentially}. We can conveniently discriminate the different
scaling behaviors by fitting $\overline{C}(r)$ in double logarithmic
representation with a function of the form
\begin{equation}
	\log \left[ \, \overline{C}(r) \right] = 
		- \alpha \, \xi^2 - \frac{K}{2} \, \xi + \mathrm{const.} \; ,
	\label{eq:hoppcorr_log}
\end{equation}
where $\xi=\log \left[ \mathrm{crd}(r) \right]$ and $\alpha$ is the curvature of the fit.
Clearly, $\alpha=0$ indicates a straight line in log-log scale and therefore an algebraic decay,
while $\alpha > 0$ captures the downward bending of the exponential decay in the same plot.
It turns out that the increase of $\alpha$ is quite sharp both in the low and high interaction regime, 
allowing us to apply this method throughout the numerically poorly explored $U < 2$ region. 
Fig.~\ref{fig:hoppcorr_averaged} demonstrates 
the technique for two different lines of constant disorder $\Delta$  
and illustrates the extraction of the estimated SF-BG transition points.
For the quantitative determination of the critical points we adopt 
a threshold for $\alpha$ equal to two standard deviations away from the average value obtained in the 
SF region. The results are summarized in Fig.~\ref{fig:pd_dirtybosons},
together with the phase boundary obtained from the Giamarchi-Schulz criterion 
in Fig.~\ref{fig:luttparam_criterion}. It is apparent that the threshold determines a phase boundary 
which is in very good agreement with the $\overline{K}_c = 2/3$ line. 

Let us notice that these results seem to indicate that the SF lobe does not shrink due to rare events,
as occurring in the mechanism leading to a ``scratched XY universality class''~\cite{Pollet2014}.
This might well be due to the system sizes and amounts of disorder realizations 
we can reasonably access with our TTN method, 
but it would pose an interesting visibility question for realistic experimental setups 
that might suffer similar limitations.
On the other hand, in the same Fig.~\ref{fig:pd_dirtybosons} the predicted critical line at tiny interactions, 
$\Delta_c \sim U^{3/4}$~\cite{Pollet2013b}, appears to be matched nicely by the prosecution 
of the phase boundary we obtained down to $U = 0.4\,$. 
In conclusion, even though the increasing computational efforts needed to explore the regime  
of very small interactions $U \ll 1$ prevent us from presenting converged and stable 
results in that region, we consider the phase diagram of Fig.~\ref{fig:pd_dirtybosons} 
representative of the physics of the system.

\section{Conclusions}
\label{conclusions}

In this manuscript we employed our recently introduced 
gauge-adaptive Tree Tensor Network algorithm~\cite{Gerster2014}
to study the disordered Bose-Hubbard model in one dimension. 
Thanks to its natural ability to deal with periodic boundary conditions, 
we have computed the superfluid stiffness,
the correlation functions and the scaling of entanglement entropy (see ~\ref{app:corr}) 
with a high degree of precision. 
We showed that criteria based on all these three quantities predict the location 
of the BKT transition for the clean (no disorder) case
within an error of a few $10^{-3}$. 

In order to identify the still partially undetermined, lobe-shaped, boundary between 
the superfluid phase and the Bose glass phase,
we considered not only disorder-averaged quantities but also their statistical distributions.
First, we showed that on the right side  of the lobe (i.e., strongly interacting regime) 
the averaged stiffness exhibits a jump in the thermodynamic limit that can be fitted with 
the conventional BKT finite-size scaling (its height being a parameter to fit).
Then, we determined the line corresponding to the averaged Luttinger parameter 
attaining a value $\overline{K}=2/3$,
and observed that this is consistent with the line that discriminates
between self-averaging and broadening regimes of its statistical distribution.
The latter was indeed recently indicated~\cite{Pollet2013a} as a more 
reliable criterion for the left side of the lobe (i.e., weakly interacting regime),
where a new universality class might take over the SF-BG transition.
Finally, we put forward a criterion based on the decay of disorder-averaged correlation functions:
the line determined by their change between algebraic and exponential decay also nicely agrees 
with the other established criteria.

The high compatibility between the results from the various analysis methods we employed
is a strong proof of reliability of our numerical study. 
In conclusion, hardly any difference is detectable 
between the ``self-averaging'' criterion and the $\overline{K}=2/3$ one 
(or alternatively our averaged correlation decay),
at least for what concerns system sizes and disorder repetitions we explored, 
which however are compatible with the conditions of current experimental verification.
Finally, we stress that our data nicely match the analytically predicted critical line 
at tiny interactions~\cite{Pollet2013b}, thus reconciling the two pictures.

As an interesting perspective, we mention here the possible extension of the TTN method to 
the time-dependent case which will allow to consider non-equilibrium situations that 
are out of reach with current numerical techniques.

\ack

We gratefully acknowledge discussions with L. Pollet, N. Prokof'ev and A. Pelster.
We acknowledge financial support from the EU integrated projects SIQS, RYSQ
and QUIC, from Italian MIUR via PRIN Project 2010LLKJBX, from the Oxford Martin 
School, from the DFG via the SFB/TRR21, and from the project OSCAR.
The numerical calculations have been performed with the computational resources provided by
the bwUniCluster~\cite{bwunicluster} in Ulm and by the MOGON Cluster of the ZDV Data Center in Mainz.
R.F. also acknowledges the Clarendon Laboratory for hospitality where part of 
the work has been performed.

\appendix

\section{Accuracy of the correlation functions and 
determination of the critical point via entanglement entropy}
\label{app:corr}

\begin{figure}
	\centering
	\includegraphics{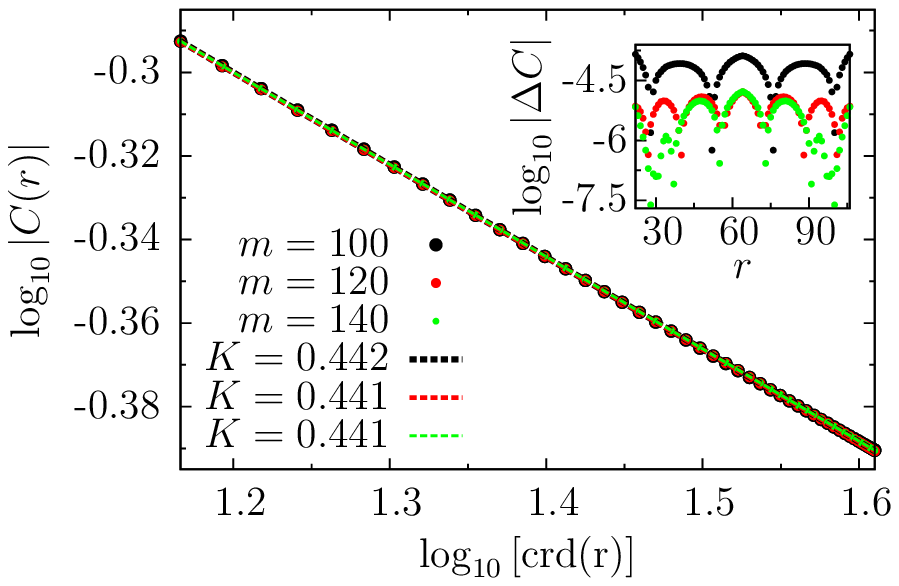}	
	\caption{\label{fig:luttparam_bonddim} Hopping correlation functions $C(r)$ and
	fitted Luttinger parameters $K$ for three different bond dimensions $m$ in the clean case $\Delta=0$. 
	The system size is $N_s=128$ and the interaction strength is $U=3.1$. 
	The accuracy of $K$ is of the order of $10^{-3}$. The inset shows 
	the difference $\Delta C$ between the data and the fitted lines.
	}
\end{figure}

In this appendix we show that the TTN state representation combined 
with the PBC setting allows for an accurate determination of critical exponents. 
For the system sizes $N_s\leq 256$ used in this work, our
computational resources allow to use bond dimensions of up to $m = 150$.
At these bond dimensions the critical exponents fitted to the correlation
functions have an accuracy of the order of $10^{-3}$, which is why no
extrapolation in $m$ is needed for our purposes. 
In Fig.~\ref{fig:luttparam_bonddim} we demonstrate this for the clean system, 
similar results are obtained for the disordered case.

\begin{figure}
	\centering
	\includegraphics{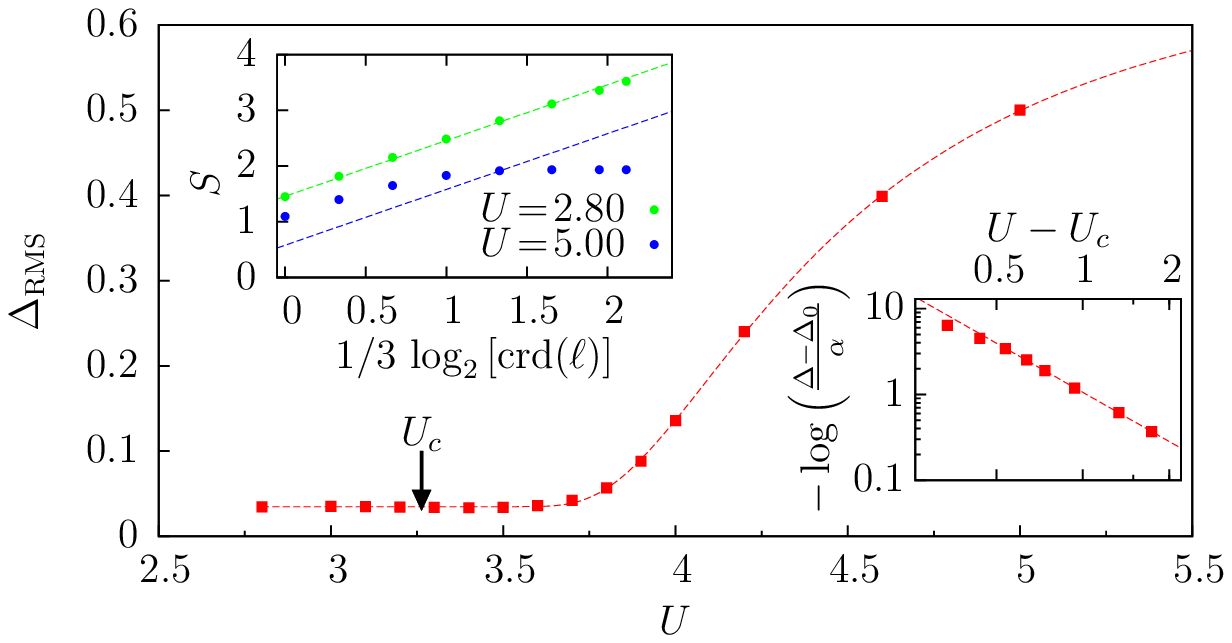}	
	\caption{\label{fig:entropy}Determination of the BKT transition point  
	by probing the scaling of the von Neumann entropy~$S$ for a ring of $N_s=256$
	sites without disorder ($\Delta=0$). After fitting
	Eq.~\eref{eq:entropy} for various values of $U$ (see left inset for two examples,
	dashed lines are fits), the resulting RMS errors are used to determine $U_c$
	via a fit of the form Eq.~\eref{eq:rms_entropy}. Extracted fit parameters are
	$U_c=3.26 \pm 0.13$, $\alpha=0.67 \pm 0.04$, $\beta = 1.9 \pm 0.4$, $\gamma=1.1 \pm 0.2$,
	$\Delta_0 = 0.035 \pm 0.001$. The right inset shows the same data as the main 
	panel, but scaled in such a way that the power-law inside the exponential of  
	Eq.~\eref{eq:rms_entropy} becomes apparent.
	}
\end{figure}

Finally, in order to complement the data on the BKT transition in the
clean system, we use the entanglement entropy
(a quantity that is easily accessible by means of tensor network 
methods~\cite{Schollwoeck2005-2011}) as an alternative criterion 
to locate the phase transition.
To this end, we numerically calculate the von Neumann entropy $S(\ell)$ for
various bipartitions $\ell\,|\,N_s-\ell$ of the ring.
Since the SF phase is known to be critical (with central charge $c=1$),
the entanglement entropy obeys \cite{Calabrese2004}
\begin{equation}
	S(\ell) =  \frac{c}{3} \log \left[ \mathrm{crd}(\ell) \right] + a \; ,
	\label{eq:entropy}
\end{equation}
with $a$ a non-universal constant. By fitting Eq.~\eref{eq:entropy} ($a$ being
the only free parameter) for different values of $U$, we expect to get an
increase in the fit's RMS error $\Delta_\mathrm{RMS}$ starting from $U_c$, where
Eq.~\eref{eq:entropy} is not valid any more because in the MI phase 
the von Neumann entropy $S(\ell)$ saturates for large $\ell$ (obeying an area law).
As shown in Fig.~\ref{fig:entropy}, we find that the behavior of
$\Delta_\mathrm{RMS}$ as a function of $U$ is well described by the form
\begin{equation}
	\Delta_\mathrm{RMS}(U) = 
	\cases{
		\Delta_0 + \alpha \, \exp \left( -\frac{\gamma}{(U-U_c)^\beta} \right)	&	for $U > U_c$ \\
		\Delta_0	&	for $U \leq U_c$ \\
	} \; ,
	\label{eq:rms_entropy}
\end{equation}
which is closely related to the opening of the energy gap 
$E_g \sim \exp \left( - \delta / \sqrt{U-U_c} \right)$ at the BKT transition
\cite{Kuehner1998}, with some constants $\Delta_0$, $\alpha$, $\beta$, $\gamma$, $\delta$.
By fitting Eq.~\eref{eq:rms_entropy} to our data we obtain a numerical value of $U_c =
3.26 \pm 0.13$ for the transition point (equivalently, $t_c = 0.306 \pm 0.011$), 
see Fig.~\ref{fig:entropy}. Despite the fact that this result is compatible with the ones 
we obtained by means of the other methods, its precision is about one order of magnitude worse.
This is due to the exponentially slow increase of $\Delta_\mathrm{RMS}$ in Eq.~\eref{eq:rms_entropy}, 
which prevents a more accurate determination of $U_c$.

In Tab.~\ref{tab:results_compare} we summarize the results for the SF-MI 
transition points $t_c$ obtained from the various techniques 
that we employed (superfluid density in Section~\ref{sssection}, 
Luttinger parameter in Section~\ref{corrsection}, entanglement entropy). 
The mutual agreement of the obtained values is of the order of some $10^{-3}$.

\begin{table}
	\caption{\label{tab:results_compare}Comparison of the critical hopping strengths
	$t_c$ for the 1D Bose-Hubbard model without disorder, obtained from various 
	numerical methods. Also shown are two values reported 
	in earlier works~\cite{Kuehner2000,Kashurnikov1996}, 
	see Refs.~\cite{Carrasquilla2013,Rachel2012,Krutitsky2015} for a comprehensive summary.}
	\begin{indented}
		\lineup
		\item[]\begin{tabular}{ll}
			\br
			Method & $t_c / U$	\\
			\mr
			Superfluid density	&	$0.295 \pm 0.003$	\\
			Luttinger parameter	&	$0.299 \pm 0.002$	\\
			Entanglement entropy	&	$0.306 \pm 0.011$ \\
			K\"uhner et al. \cite{Kuehner2000}	&	$0.297 \pm 0.010$	\\
			Kashurnikov and Svistunov \cite{Kashurnikov1996}	&	$0.304 \pm 0.002$	\\
			\br
		\end{tabular}
	\end{indented}
\end{table}

\section{Convergence of distributions with respect to the number of disorder realizations}
\label{app:sample_number}

\begin{figure}
	\centering
	\includegraphics{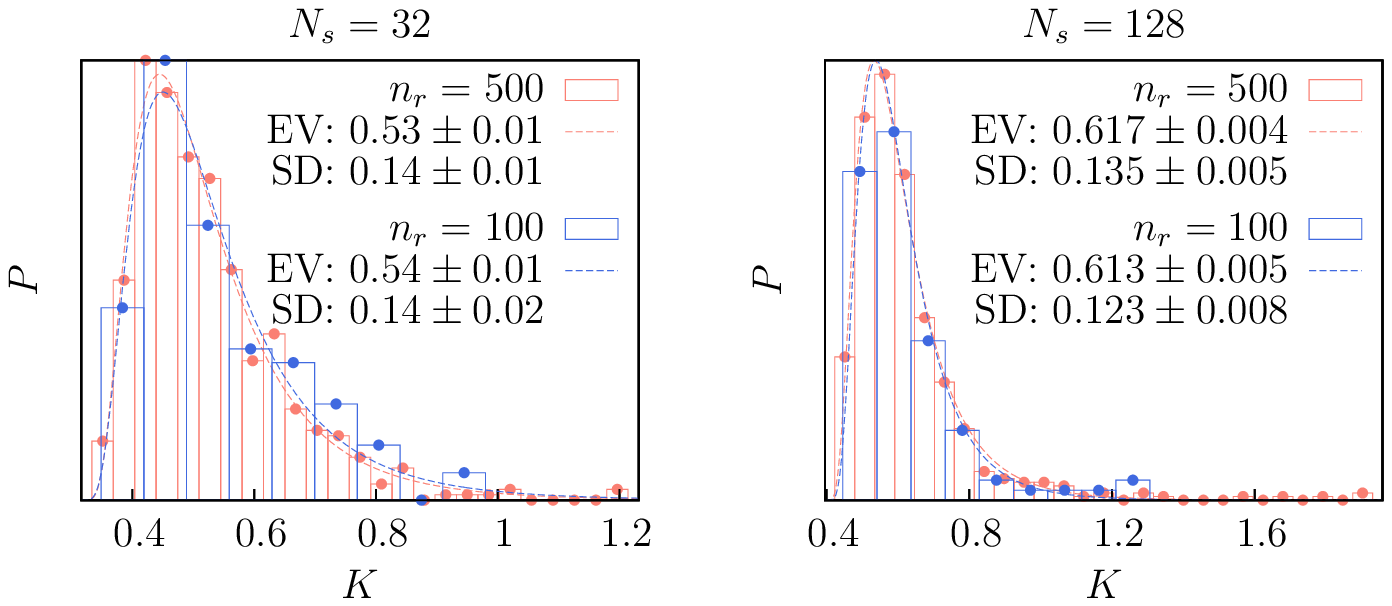}
	\caption{\label{fig:sample_number}Distributions of the Luttinger parameter $K$
	for two different system sizes~$N_s$ and two different numbers of disorder
	realizations $n_r$. In both plots the disorder strength is $\Delta=3.0$ and 
	the interaction is $U=2.2$. The characteristics of the distributions 
	(in particular their expectation value (EV) and 
	standard deviation (SD)) remain within the statistical error  
	when increasing $n_r$ from $100$ to $500$. 
	Fitted lines are log-normal distributions.}
\end{figure}

In Fig.~\ref{fig:sample_number} we demonstrate that the distributions of the
Luttinger parameter $K$ can be meaningfully characterized with a number of
disorder realizations of the order of $n_r=100$, even in the regime of relatively strong disorder $\Delta$.
This statement holds true for all system sizes $N_s$ used in this work. 
Moreover, it becomes apparent that the distributions can be faithfully described by the log-normal 
distributions defined in Eq.~\eref{eq:lognormal}. 
We remark that the identification of rare events with very large $K$, 
as described in~\cite{Pollet2013a,Pollet2013b}, would require system sizes and number of 
disorder realizations of the order of  several thousand, which is not within  
the scope of the present investigation (see main text).

\section{Comparison of phase diagram with data from the literature}
\label{app:phasediag_comp}

\begin{figure}
	\centering
	\includegraphics{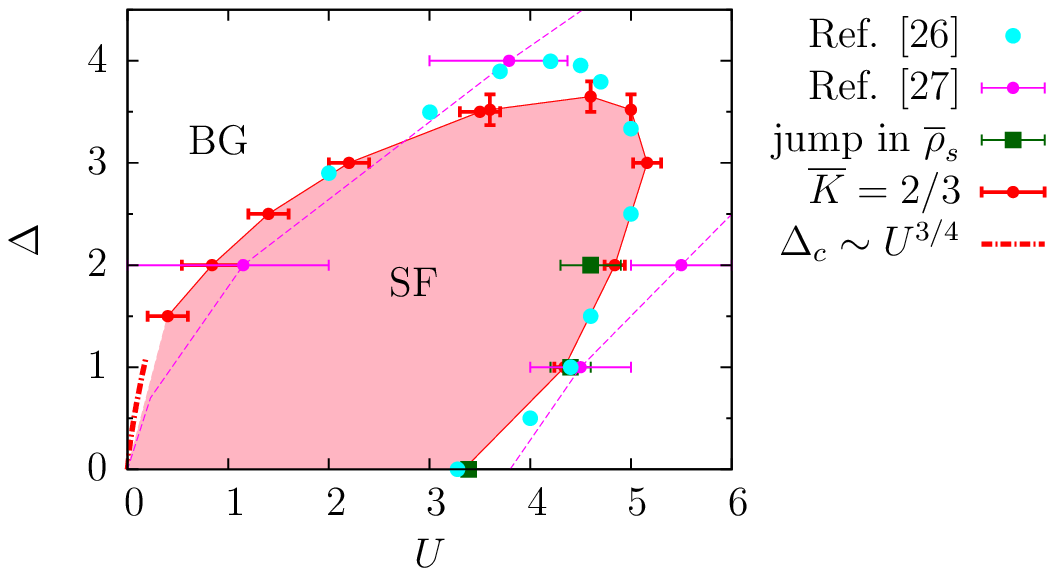}
	\caption{\label{fig:pd_comp}Comparison of the phase diagram from 
	Fig.~\ref{fig:pd_dirtybosons} with phase diagrams previously 
	obtained in the literature.}
\end{figure}

In this appendix we compare our data on the extension of the superfluid lobe  
(as summarized in Fig.~\ref{fig:pd_dirtybosons}) with results previously obtained in the literature, 
in particular with data from Prokof'ev \emph{et al.}~\cite{Prokofev1998} and 
Rapsch \emph{et al.}~\cite{Rapsch1999}. A direct comparison is 
shown in Fig.~\ref{fig:pd_comp}. As already mentioned in the main text, 
we observe good agreement with the Monte Carlo results reported in Ref.~\cite{Prokofev1998} 
(except for a slight deviation at the tip of the lobe).

\section{Local boson occupation numbers}
\label{app:local_occupations}

\begin{figure}
	\centering
	\includegraphics{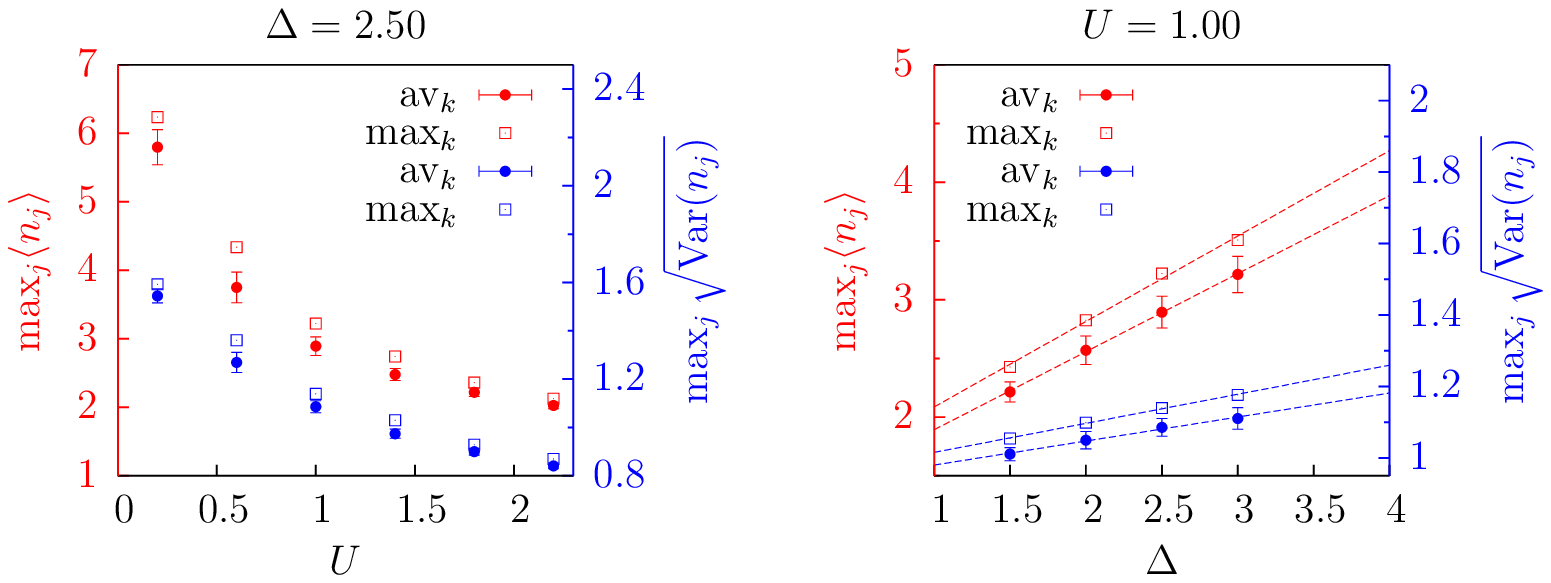}
	\caption{\label{fig:local_occup}
	Local boson occupation expectation value and variance 
	as a function of $U$ (left) and $\Delta$ (right).
	Red filled circles: realization-averaged maximally occurring expectation value 
	$\mathrm{av}_k [ \max_j \langle n_j \rangle ]$ (indices $k$ and $j$ run over  
	$n_r=100$ disorder realizations and $N_s=128$ sites, respectively). 
	Red squares: highest measured expectation value 
	$\max_k [ \max_j \langle n_j \rangle ]$. 
	Blue filled circles: realization-averaged maximally occurring standard deviation  
	$\mathrm{av}_k [ \max_j \sqrt{ \mathrm{Var}(n_j) } ]$.
	Blue squares: highest measured standard deviation 
	$\max_k [ \max_j \sqrt{ \mathrm{Var}(n_j) } ]$.	
	The plotted error bars are standard deviations determined from the realization ensembles. 
	Dashed lines are linear fits.}
\end{figure}

In order to get an estimate for the necessary local (on-site) dimension $d$, one 
has to consider the expectation values $\langle n_j \rangle$ of the local boson 
number operators and their corresponding variances 
$\mathrm{Var}(n_j) = \langle n_j^2 \rangle - \langle n_j \rangle^2$. We plot these
quantities in Fig.~\ref{fig:local_occup}, both as a function of interaction 
strength $U$ (with disorder strength $\Delta$ fixed) and as a function 
of $\Delta$ ($U$ fixed). Adding two standard deviations to the highest occurring
boson occupation expectation values along the ring, we find that up to $d=10$ 
states are relevant for the lowest $U$ values we simulated. By choosing $d=20$ for
this regime (as reported in the main text) we can be absolutely sure 
that errors due to the truncation of the local boson bases are negligible. We 
can easily afford such high local dimensions~$d$ because the impact on the 
computation time is relatively small in TTN simulations (e.g. for a 
$N_s=128$ simulation the runtime only increases by a factor of~$1.3$ when 
using $d=20$ instead of $d=10$). 

Furthermore, Fig.~\ref{fig:local_occup} suggests 
that for constant $U$ the maximally occurring occupation number increases 
linearly with the disorder strength~$\Delta$.

\end{document}